\documentclass[conference]{IEEEtran}
\makeatletter
\newcommand\fs@betterruled{%
  \def\@fs@cfont{\bfseries}\let\@fs@capt\floatc@ruled
  \def\@fs@pre{\vspace*{5pt}\hrule height.8pt depth0pt \kern2pt}%
  \def\@fs@post{\kern2pt\hrule\relax}%
  \def\@fs@mid{\kern2pt\hrule\kern2pt}%
  \let\@fs@iftopcapt\iftrue}

\IEEEoverridecommandlockouts
\usepackage{cite}
\usepackage{float}
\usepackage{amsmath,amssymb,amsfonts}
\usepackage{graphicx}
\usepackage{textcomp}
\usepackage{xcolor}
\usepackage{enumitem}
\usepackage{subcaption}
\usepackage{multirow}
\usepackage{algorithm}
\usepackage{algpseudocode}
\usepackage{hyperref}

\usepackage[textwidth=30mm]{todonotes}
\usepackage{soul}

\setstcolor{magenta}
\sethlcolor{lightgray}

\bstctlcite{BSTcontrol}
\begin{document}

\title{On the Uplink Performance of Finite-Capacity Radio Stripes}

\author{
\IEEEauthorblockN{Ioannis Chiotis and Aris L. Moustakas}
\IEEEauthorblockA{Physics Dept.,
National Kapodistrian University of Athens, Zografou, Greece}
Email: \href{mailto:ioachiotis@phys.uoa.gr}{\{ioachiotis}, \href{mailto:arislm@phys.uoa.gr}{arislm}\}@phys.uoa.gr
}

\maketitle

\begin{abstract}
Cell-Free (CF) Massive MIMO (mMIMO) is a technology which can potentially augment not only the deployment of 5G, but also the deployment of beyond 5G (B5G) wireless networks. However, the cost for rolling out such systems may be significant. Radio stripes form a promising solution which offers the potential of scalability at a reduced price. This paper investigates the uplink scenario of a CF mMIMO system, implemented with a limited-capacity radio stripe which integrates a novel arrangement of access points (APs), fully exploiting macro-diversity benefits. We also analyze a heuristic Compare-and-Forward (CnF) strategy, which, by comparing normalized linear minimum mean square error (N-LMMSE) soft estimates, enables optimal dynamic cooperation clustering, thus leading to a user-centric radio stripe network approach. Aiming at maximizing the per-user uplink spectral efficiency (SE), we ensure that, under finite capacity constraints, our solution can guarantee better performance than existing radio stripe architectures, especially when system size scales.
\end{abstract}


\section{Introduction}
Since the appearance of CF mMIMO systems, a significant segment of the literature research has focused on the enhancement of star-like solutions, which assume direct connections between the centralized processing unit (CPU) and the APs, since they lead to higher SE \cite{versus, making} and energy efficiency (EE) \cite{total, energy} benefits compared to conventional small cells. However, these individual links not only require significant expenses in order to be deployed and maintained, but also lead to system rigidness, not allowing for easy relocation of currently deployed equipment as user-location dynamically changes. Moreover, as service demands keep on growing, centralized layouts will soon exhaust their computational capabilities. Therefore, there is an urgent need to turn to more distributed solutions in order to route contemporary issues as well as to prepare for the B5G challenges.

An attractive solution, which could potentially deal with the aforementioned issues, is this of the radio stripe architecture \cite{stripes, ubiquitous, patent, qlmmse}. An interesting sequential processing algorithm that can truly exploit the serial connection of the elements on a radio stripe and which motivated this study is presented in \cite{stripes}. In that work it is shown that radio stripes, using that algorithm alongside with normalized linear minimum mean square error (N-LMMSE) processing, can outperform L2\cite{making} maximal ratio combing (MRC), while achieving comparable performance to the optimal L4 MMSE implementation\cite{making}. Another great radio stripe processing solution is quasi-LMMSE (Q-LMMSE) \cite{qlmmse}, which promises not only to address latency and FH complexity, but also, provided that payload period is large enough, to offer better performance than the serial N-LMMSE. However, as promising as all these may sound, radio stripes are far apart from being materialized due to finite-capacity limitations, which, especially in large-scale scenarios, can cause serious bottleneck issues as users' data are too much for one link to handle.

\textbf{Contributions:} Main purpose of this paper is to study the impact that a finite-capacity radio stripe has on the per-user uplink throughput, when used as a FH network of a CF mMIMO topology, especially when system extent scales. Furthermore, we propose a potent CnF strategy which can effectively assign a dynamic cooperation cluster (DCC) \cite{beyond, anewlook} to each one of the user equipments (UEs) the network serves, focusing on their SINR maximization. That strategy leads to a novel user-centric radio stripe approach that enables the sequential N-LMMSE\footnote{More optimized Q-LMMSE design is avoided because its superior performance is only valid for scenarios with large payload period, a fact that does not match to rapidly varying environments with extensive demands for service. Additionally, since our purpose is to test the system as it scales, that technique would actually perform worse than sequential N-LMMSE.} combining to be performed only in antenna processing units (APUs) that can strongly contribute to each user's $k$ message signal estimation, compensating for the extra quantization error that will be attached to their soft estimation upon its retransmission into the stripe. That way redundant analog-to-digital conversions are avoided, a fact that is rather crucial, especially in extensive setups. In addition, we introduce a heuristic radio stripe arrangement of fully distributed APs that further enhances performance, taking advantage of macro-diversity. However, since these distributed APs are attached to the main body of the stripe (and not directly to the APUs), they also occupy portion of the total available rate. Hence, a simple rate allocation is also employed to efficiently share the radio stripe's capacity between its components.

\begin{figure}[ht]
\begin{subfigure}{.5\textwidth}
  \centering
  \includegraphics[width=.70\linewidth]{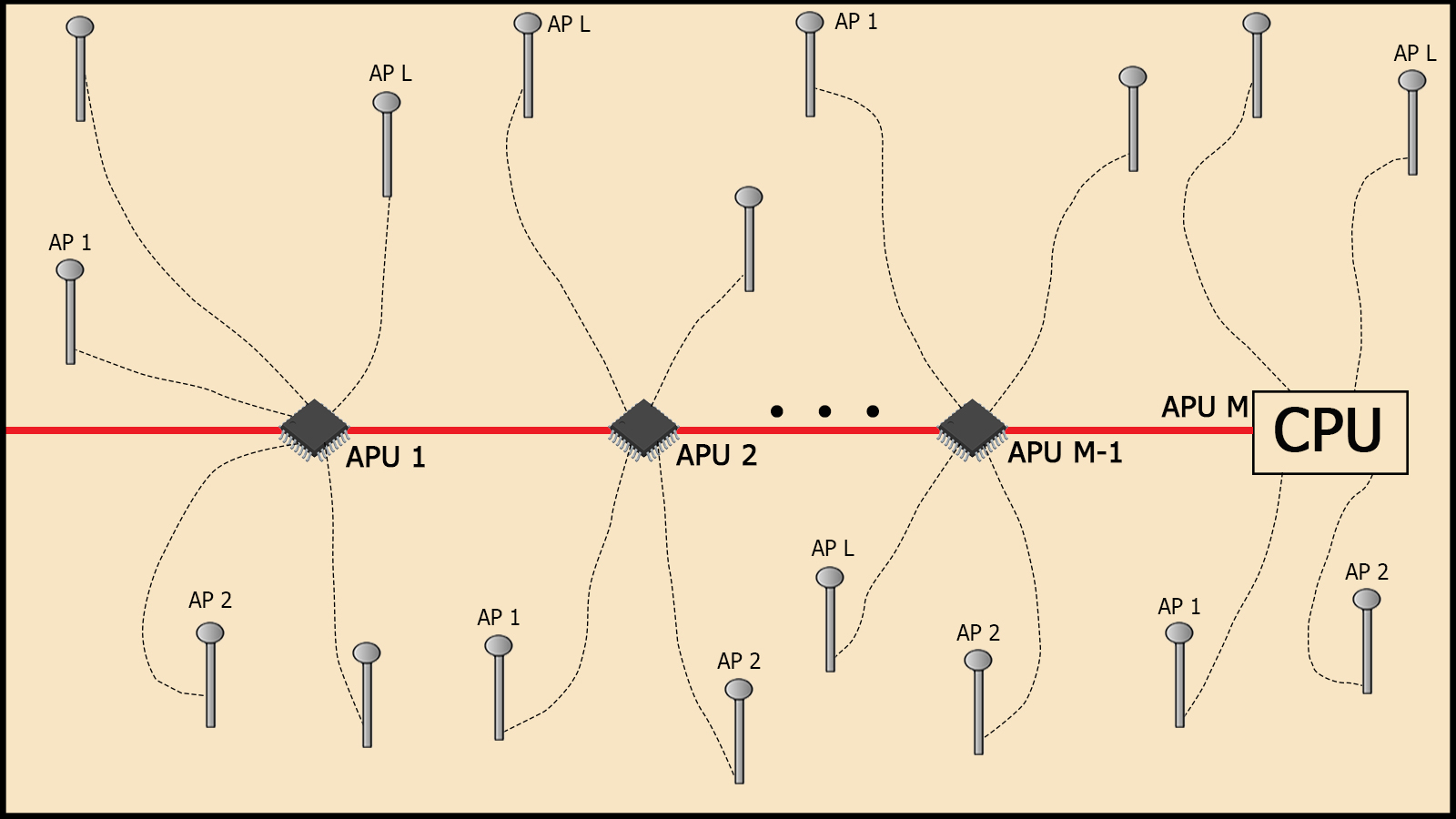}  
  \caption{Radio stripe with non-integrated APs on the film.}\hfill
  \label{2a}
\end{subfigure}
\begin{subfigure}{.5\textwidth}
  \centering
  \includegraphics[width=.70\linewidth]{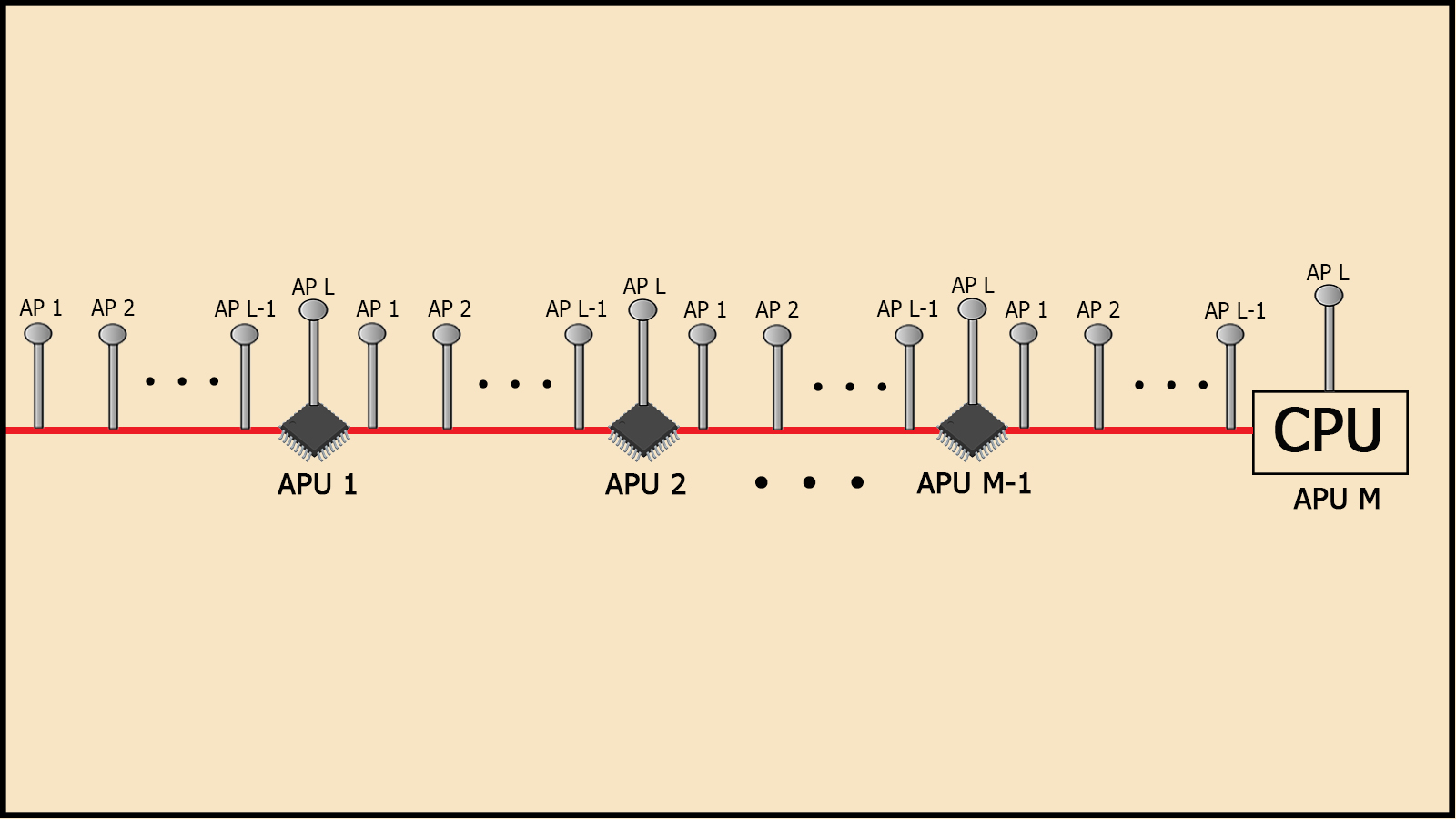}
  \caption{Radio stripe with integrated APs on the film.}
\label{2b}
\end{subfigure}
\caption{Proposed radio stripe scheme.}
\label{2}
\end{figure}

Radio stripe topology has been based on the idea that several APUs, each with a number of $N$ co-located APs \cite{stripes, qlmmse}, are connected through an infinite-capacity bus, interchanging data sequentially. To enhance that layout, we propose a new scheme that even the APs follow a distributed setup, as it is depicted in Fig. \ref{2a}. In addition, for practical purposes, we also suggest the integration of the distributed APs of each APU into the bus itself, resulting to the setup in Fig. \ref{2b}. Assuming a simple orthogonal resource allocation scheme, e.g. frequency-division multiplexing (FDM), each element of the radio stripe uses its allocated bandwidth to transmit its data. As a result, the bus connecting any two APUs is used to transfer not only inter-APU data, but also pilot and data signals from the APs located on that part of the bus to the APU on the right. Of course, the sum rate of all the components in each part between two successive APUs is constrained to the total capacity of the radio stripe. 

\textit{\textbf{Notations:}} Superscripts $(.)^{*}$, $(.)^\dagger$, $(.)^T$ and $(.)^{-1}$ are conjugate, Hermitian transpose, transpose and inverse operators respectively. Bold uppercase letters (e.g. $\mathbf{B}$) denote matrices and bold lowercase letters (e.g. $\mathbf{v}$) column vectors. $\mathbf{0}_L$ signifies a vertical zero vector of $L$ dimensions, $\mathbf{O}_L$ a zero matrix of $L\times L$ dimensions and $\mathbf{I}_{L}$ an identity matrix of the same dimensions. Circularly-symmetric variables that follow complex normal distribution with correlation matrix $\mathbf{R}$ and zero mean are denoted as $\mathcal{CN}(\mathbf{0},\mathbf{R})$, while expected value and variance as $\mathbb{E}\{.\}$ and $\text{Var}\{.\}$ respectively. Finally, $|.|$, $||.||$, $\triangleq$ and $\oplus$ symbolize the absolute value of a scalar, $l_2$ norm of a vector, definitions and direct sum of matrices respectively.

\section{Radio Stripe Channel Model}
We consider the uplink scenario of the architecture in Fig. \ref{2b}, comprising of $M$ on-film integrated APUs (the $M^{th}$ is the CPU), each connected with $L$ on-film distributed single-antenna APs. In the layout, we consider $K$ single-antenna UEs, where the channel between UE $k$ and APU $m$ is denoted by $\mathbf{g}_{mk}\in\mathbb{C}^{L}$, where $m\in[1,M]$ and $k\in[1,K]$. The channel is assumed to be constant over a coherence interval $\tau_{c}$ and it is drawn from a Rayleigh fading distribution as:
\begin{equation}\label{eq:gmk}
  \mathbf{g}_{mk}\sim\mathcal{CN}(\mathbf{0},\mathbf{R}_{mk})
\end{equation}
where $\mathbf{R}_{mk}\in\mathbb{C}^{L\times L}$ represents the spatial correlation matrix, which is assumed to be diagonal and known. Its diagonal elements $\mathbf{\beta}_{mkl}$, for $l\in[1,L]$, form the large-scale path loss between antenna $l$ and UE $k$. 

We also assume that uplink training period lasts for $\tau_{p}$ samples, where $\tau_{p}<\tau_{c}$, in which all UEs simultaneously transmit mutually orthogonal $\tau_{p}$-length pilot signals $\sqrt{\tau_{p}}\boldsymbol{\phi}_{k}\in\mathbb{C}^{\tau_{p}}$, with ${\Vert\boldsymbol{\phi}_{k}\|}^2=1$. For simplicity, we neglect pilot contamination, hence $\tau_{p}\geq K$. Quantized pilot matrix $\mathbf{\bar{Y}}_{p,m}\in\mathbb{C}^{L\times \tau_{p}}$ received by APU $m$ during that phase equals to:
\begin{equation}\label{eq:barYpm}
  \bar{\mathbf{Y}}_{p,m}=\sqrt{\rho\tau_{p}}\sum_{i=1}^{K}\mathbf{g}_{mi}\boldsymbol{\phi}_{i}^T+\mathbf{N}_m+\mathbf{E}^{\mathbf{Y}_{p,m}}
\end{equation}
where $\rho\geq0$ is the transmit power of each UE, $\mathbf{N}_m$ is the noise matrix of each APU $m$ and $\mathbf{E}^{\mathbf{Y}_{p,m}}$ is the quantization error matrix due to the finite-capacity constraint, both assumed to contain i.i.d. complex Gaussian elements with zero mean and variance $\sigma^2_n$ and $\sigma^2_{e,p_{\bar{m}}}$\footnote{As will be shown in Section VI, diagonal elements of $\text{Var}\{\mathbf{E}^{\mathbf{Y}_{p,m}}\}$, $\forall m\in[1,K]$, differ. For computational simplicity, in each APU $m$, we use the average $\sigma^2_{e,p_{\bar{m}}}$ that has been taken over all of its $L$ APs. The same also applies to $\sigma^2_{e,\bar{y}}$, which is additionally averaged over all $M$ APUs. That helps us manage $\sigma^2_{e,\bar{y}}$ the same way we manage thermal noise $\sigma^2_n$.} respectively. Then, according to \cite[Chap. 3]{emil}, the MMSE channel estimate $\hat{\mathbf{g}}_{mk}\in\mathbb{C}^{L}$ is given by:
\begin{equation}\label{eq:hatgmk}
  \hat{\mathbf{g}}_{mk}=\sqrt{\rho\tau_p}\mathbf{R}_{mk}\boldsymbol{\Psi}_{mk}^{-1}\check{\mathbf{y}}_{p,m}
\end{equation}
where
\begin{equation}\label{eq:checkypm}
  \check{\mathbf{y}}_{p,m}=\bar{\mathbf{Y}}_{p,m}\boldsymbol{\phi}_{k}^*=\sqrt{\rho\tau_{p}}\mathbf{g}_{mk}+\mathbf{N}_{m}\boldsymbol{\phi}_{k}^*+\mathbf{E}^{\mathbf{y}_{p,m}}\boldsymbol{\phi}_{k}^*
\end{equation}
\begin{equation}\label{eq:Psimk}
 \boldsymbol{\Psi}_{mk}=\mathbb{E}\{\check{\mathbf{y}}_{p,m}\check{\mathbf{y}}^{\dagger}_{p,m}\}=\rho\tau_p\mathbf{R}_{mk}+\mathbf{I}_{L}(\sigma^2_n+\sigma^2_{e,p_{\bar{m}}})
\end{equation}
Note that since real channels $\mathbf{g}_{mk}$ differ from the estimated $\hat{\mathbf{g}}_{mk}$ ones, an independent estimation error $\tilde{\mathbf{g}}_{mk}=\mathbf{g}_{mk}-\mathbf{\hat{g}}_{mk}$ occurs, which is assumed to be distributed as $\mathcal{CN}(\mathbf{0},\tilde{\mathbf{R}}_{mk})$, where $\tilde{\mathbf{R}}_{mk}$ is given by:
\begin{equation}\label{eq:errorcov}
  \tilde{\mathbf{R}}_{mk}=\mathbf{R}_{mk}-\rho\tau_p\mathbf{R}^2_{mk}\boldsymbol{\Psi}_{mk}^{-1}
\end{equation}\par

During uplink payload period, all $K$ users simultaneously transmit their data to the APs for a duration of $\tau_d=\tau_c-\tau_p$ slots. The instantaneous quantized signal $\bar{\mathbf{y}}_m\in\mathbb{C}^L$, that each APU $m$ receives in each slot, is given by:
\begin{equation}\label{eq:qdata}
  \mathbf{\bar{y}}_m=\sqrt{\rho}\sum_{i=1}^K\mathbf{g}_{mi}q_i+\mathbf{n}_m+\mathbf{e}_{y_m}
\end{equation}
where $q_i\sim\mathcal{CN}(0,1)$ is the transmitted message signal from UE $i$ and $\mathbf{e}_{y_m}\sim\mathcal{CN}(0,\mathbf{I}_L\sigma^2_{e,\bar{y}})$ is the quantization error vector owed to $\mathbf{y}_m$ digitalization. Notice that the elements of the noise vector $\mathbf{n}_{m}$ have the same distribution as those of $\mathbf{N}_m$.

\section{Serial Processing Analysis}
In this section we describe the serial algorithm of \cite{stripes} taking also into account the finite-capacity constraints, which impose a quantization error, with zero mean and variance specified by the corresponding rate, to be attached on every signal transmitted through the radio stripe. For simplicity, in the present paper, we only analyze the case where side information is exchanged between APUs without any distortion.

Firstly, APU 1, in order to estimate message $q_k$, $\forall k\in[1,K]$, it designs the combining vector $\left\{\mathbf{v}_{1_k}:k\in[1,K]\right\}\in\mathbb{C}^{L}$, where $||\mathbf{v}_{1_k}||^2=1$, using solely local information $\left\{\Theta_{1_k}=\{\mathbf{\hat{g}}_{1i},\mathbf{\tilde{R}}_{1i}\}:k,i\in[1,K]\right\}$. Then, soft estimate $\hat{s}_{1_k}$ of $q_k$ arises by associating this vector with the one in \eqref{eq:qdata} as:
\begin{equation}\label{eq:hats1k}
\begin{split}
  &\hat{s}_{1_k}=\mathbf{v}^{\dagger}_{1_k}\mathbf{\bar{y}}_1=\sqrt{\rho}\sum_{i=1}^Kz_{1_ki}q_i+n_{1_k}+e^{y}_{1_k}
    \end{split}
\end{equation}
where $z_{1_ki}\triangleq\mathbf{v}^{\dagger}_{1_k}\mathbf{g}_{1i}$, $n_{1_k}\triangleq\mathbf{v}^{\dagger}_{1_k}\mathbf{n}_1$ and $e^{y}_{1_k}\triangleq\mathbf{v}^{\dagger}_{1_k}\mathbf{e}_{y_1}$. Also, note that $n_{1_k}\sim\mathcal{CN}(0,\sigma^2_n)$ and $e^{y}_{1_k}\sim\mathcal{CN}(0,\sigma^2_{e,\bar{y}})$, since $\mathbf{v}_{1_k}$ is normalized\cite{stripes}. Afterwards, in order for APU 2 to further enhance $\hat{s}_{1_k}$ \cite{stripes}, APU 1, based on $\{\Theta_{1_k}\}$, shares with it all the above calculated soft estimates $\hat{s}_{1_k}$, as well as the effective channel estimates $\hat{z}_{1_ki}\triangleq\mathbf{v}^{\dagger}_{1_k}\mathbf{\hat{g}}_{1i}$ and the sum of all the effective channel error variances $\tilde{\psi}_{1_k}\triangleq\sum_{i=1}^K\tilde{\psi}_{1_ki}$ as side information\footnote{In contrast to \cite{stripes}, we assume that each APU uses as side information the sum of all the effective error variances, aiming at the traffic reduction in the FH. Note that using $\tilde{\psi}_{1_k}$ instead of $\tilde{\psi}_{1_ki}$, $\forall k,i\in[1,K]$, helps to reduce data traffic by a factor of $K$.}, since channels $\mathbf{g}_{1k}$ are unknown to it. Note that $\tilde{\psi}_{1_ki}$ arises as $\left\{\tilde{z}_{1_ki}\triangleq\mathbf{v}^{\dagger}_{1_k}\mathbf{\tilde{g}}_{1i}\right\}\sim\mathcal{CN}(0,\tilde{\psi}_{1_ki})$ and equals to:
\begin{equation}\label{eq:tildepsi1ki}
  \tilde{\psi}_{1_ki}\triangleq\mathbf{v}^{\dagger}_{1_k}\mathbf{\tilde{R}}_{1i}\mathbf{v}_{1_k}
\end{equation}
Due to the finite-capacity constraint, APU 2 receives:
\begin{enumerate}
    \item quantized soft estimates $\left\{\bar{s}_{1_k}=\hat{s}_{1_k}+e_{\hat{s}_{1_k}}:k\in[1,K]\right\}$
    \item side information $\left\{\delta_{1_k}=\{\hat{z}_{1_ki},\tilde{\psi}_{1_k}\}:i,k\in[1,K]\right\}$
\end{enumerate}
where $e_{\hat{s}_{1_k}}\sim\mathcal{CN}(0,\sigma^2_{e,\hat{s}_{1_k}})$. 

Then, using \eqref{eq:qdata} and \eqref{eq:hats1k}, APU 2 creates an augmented receive vector as:
\begin{equation}\label{eq:recaug}
\begin{split}
\begin{bmatrix}\mathbf{\bar{y}}_2 \\ \bar{s}_{1_k} \end{bmatrix}=\sqrt{\rho}\sum_{i=1}^K\begin{bmatrix}\mathbf{g}_{2i} \\ z_{1_ki} \end{bmatrix}q_i+\begin{bmatrix}\mathbf{n}_2 \\ n_{1_k} \end{bmatrix}+\begin{bmatrix}\mathbf{e}_{y_2} \\e^{y}_{1_k} \end{bmatrix}+\begin{bmatrix}\mathbf{0}_L \\ e_{\hat{s}_{1_k}} \end{bmatrix}
    \end{split}
\end{equation}

Next, using $\left\{\Theta_{2_k}=\{\mathbf{\hat{g}}_{2i},\mathbf{\tilde{R}}_{2i},\delta_{1_k}\}:i,k\in[1,K]\right\}$ as side\linebreak information, APU 2 designs an augmented combining vector $\left\{\mathbf{v}_{2_k}:k\in[1,K]\right\}\in\mathbb{C}^{L+1}$, where $||\mathbf{v}_{2_k}||^2=1$. Then, as in \eqref{eq:hats1k}, augmented soft estimate $\hat{s}_{2_k}$ is expressed as:
\begin{equation}\label{eq:augsoftestim}
\begin{split}
  \hat{s}_{2_k}=\mathbf{v}^{\dagger}_{2_k}\begin{bmatrix}\mathbf{\bar{y}}_2 \\ \bar{s}_{1_k} \end{bmatrix}=\sqrt{\rho}\sum_{i=1}^K z_{2_ki}q_i+n_{2_k}+e^{y}_{2_k}+e^{s}_{2_k}
    \end{split}
\end{equation}
where $z_{2_ki}\triangleq\mathbf{v}^{\dagger}_{2_k}\begin{bmatrix}\mathbf{g}_{2i} \\ z_{1_ki} \end{bmatrix}$, $n_{2_k}\triangleq\mathbf{v}^{\dagger}_{2_k}\begin{bmatrix}\mathbf{n}_2 \\ n_{1_k}\end{bmatrix}$, $e^{y}_{2_k}\triangleq\mathbf{v}^{\dagger}_{2_k}\begin{bmatrix}\mathbf{e}_{y_2} \\ e^{y}_{1_k}\end{bmatrix}$ and $e^{s}_{2_k}\triangleq\mathbf{v}^{\dagger}_{2_k}\begin{bmatrix}\mathbf{0}_L \\ e_{\hat{s}_{1_k}} \end{bmatrix}$. In addition, $e^{s}_{2_k}\sim\mathcal{CN}(0,\sigma^2_{e^{s},2_k})$, while $n_{2_k}$ and $e^y_{2_k}$ have the same distributions as the elements of $\mathbf{n}_m$ and $\mathbf{e}_{y_m}$ in \eqref{eq:qdata} respectively. In a similar way to APU 1, side information $\left\{\delta_{2_k}=\{\hat{z}_{2_ki},\tilde{\psi}_{2_k}\}:i,k\in[1,K]\right\}$ that will accompany $\hat{s}_{2_k}$ is then created based on $\{\Theta_{2_k}\}$. Note that $\hat{z}_{2_ki}\triangleq\mathbf{v}^{\dagger}_{2_k}\begin{bmatrix}\mathbf{\hat{g}}_{2i} \\ \hat{z}_{1_ki} \end{bmatrix}$, $\left\{\tilde{z}_{2_ki}\triangleq\mathbf{v}^{\dagger}_{2_k}\begin{bmatrix}\mathbf{\tilde{g}}_{2i} \\ \tilde{z}_{1_ki} \end{bmatrix}\right\}\sim\mathcal{CN}(0,\tilde{\psi}_{2_ki})$ and $\tilde{\psi}_{2_k}\triangleq\sum_{i=1}^K\tilde{\psi}_{2_ki}=\mathbf{v}^{\dagger}_{2_k}\mathbf{\tilde{J}}_{2_k}\mathbf{v}_{2_k}$, where $\mathbf{\tilde{J}}_{2_k}$ equals\cite{stripes} to:
\begin{equation}\label{eq:augsoftestervar}
\begin{split}
  \mathbf{\tilde{J}}_{2_k}=\sum_{i=1}^K\mathbf{\tilde{R}}_{2i}\oplus\tilde{\psi}_{1_k}
    \end{split}
\end{equation}

A key difference from \cite{stripes} is that APU 2 will not necessarily transmit to APU 3 $\left\{\hat{s}_{2_k},\delta_{2_k}\right\}$ for every UE $k$. Instead, if the instantaneous mean-squared error (MSE) $|\sqrt\rho q_k-\bar{s}_{2_k}|^2$ is larger than the $|\sqrt\rho q_k-\bar{s}_{1_k}|^2$, meaning that the contribution of $\hat{s}_{2_k}$ to the estimation of $q_k$ does not outweigh the extra quantization error $e_{\hat{s}_{2_k}}$ which will be attached to it upon its transmission into the stripe, APU 2 has also the option to forward $\left\{\bar{s}_{1_k},\delta_{1_k}\right\}$. In that case, APU 2 acts as relay and no additional quantization error attaches to $\bar{s}_{1k}$, since it has already been compressed. More details regarding this decision process (CnF strategy) will be provided in the next Section. In general, APU $m$, for $m\in[2,M]$, can estimate message signal $q_k$ of user $k$ using two sources. The first is the signal received from its own $L$ APs, while the second corresponds to the signal arriving from the previous APU $m-1$, but was quantized at APU $c_k$, where $1\leq c_k<m$. In this context, the general augmented reception signal that APU $m$ has available is given as follows:
\begin{equation}\label{eq:randrecaug}
\begin{split}
  \begin{bmatrix}\mathbf{\bar{y}}_m \\ \bar{s}_{c_k} \end{bmatrix}=\sqrt{\rho}\sum_{i=1}^K\begin{bmatrix}\mathbf{g}_{mi} \\ z_{c_ki} \end{bmatrix}q_i+\begin{bmatrix}\mathbf{n}_m \\ n_{c_k}\end{bmatrix}+\begin{bmatrix}\mathbf{e}_{y_m} \\ e^{y}_{c_k}\end{bmatrix}+\begin{bmatrix}\mathbf{0}_L \\ e^{s}_{c_k}+e_{\hat{s}_{c_k}} \end{bmatrix}
    \end{split}
\end{equation}
where $e^{s}_{c_k}\sim\mathcal{CN}(0,\sigma^{2}_{e^{s},c_k})$ is the total compression error attached to the transmitted soft estimates until APU $c_k$. 
Then, using $\left\{\Theta_{m_k}=\{\mathbf{\hat{g}}_{mi},\mathbf{\tilde{R}}_{mi},\delta_{c_k}\}:i,k\in[1,K]\right\}$ information, APU $m$ designs the augmented combining vector $\left\{\mathbf{v}_{m_k}:k\in[1,K]\right\}\in\mathbb{C}^{L+1}$, where $||\mathbf{v}_{m_k}||^2=1$, which combines with the signal in \eqref{eq:randrecaug} to obtain $\hat{s}_{m_k}$ as follows:
\begin{equation}\label{eq:ckaugsoftest}
\begin{split}
  \hat{s}_{m_k}=\sqrt{\rho}\sum_{i=1}^K(\hat{z}_{m_ki}+\tilde{z}_{m_ki})q_i+n_{m_k}+e^{y}_{m_k}+e^{s}_{m_k}
\end{split}
\end{equation}
where $\hat{z}_{m_ki}\triangleq\mathbf{v}^{\dagger}_{m_k}\begin{bmatrix}\mathbf{\hat{g}}_{mi} \\ \hat{z}_{c_ki} \end{bmatrix}$, $n_{m_k}\triangleq\mathbf{v}^{\dagger}_{m_k}\begin{bmatrix}\mathbf{n}_m \\ n_{c_k}\end{bmatrix}$, $e^{y}_{m_k}\triangleq$ \makebox[\linewidth][l]{$\mathbf{v}^{\dagger}_{m_k}\begin{bmatrix}\mathbf{e}_{y_m} \\ e^{y}_{c_k}\end{bmatrix}$, $e^{s}_{m_k}\triangleq\mathbf{v}^{\dagger}_{m_k}\begin{bmatrix}\mathbf{0}_L \\ e^{s}_{c_k}+e_{\hat{s}_{c_k}} \end{bmatrix}$ and $\tilde{z}_{m_ki}\triangleq\mathbf{v}^{\dagger}_{m_k}\begin{bmatrix}\mathbf{\tilde{g}}_{mi} \\ \tilde{z}_{c_ki} \end{bmatrix}$.} In addition, note that $e^{s}_{m_k}\sim\mathcal{CN}(0,\sigma^2_{e^{s},m_k})$, while $n_{m_k}$ and $e^y_{m_k}$ are distributed as $n_{2_k}$ and $e^y_{2_k}$ respectively, since all combining vectors until APU $m$ are normalized. Finally, in case APU $m$ decides to transmit the locally created $\hat{s}_{m_k}$, the side information that will accompany it will be $\left\{\delta_{m_k}=\{\hat{z}_{m_ki}, \tilde{\psi}_{m_k}\}\mathbf|\{\Theta_{m_k}\}:i,k\in[1,K]\right\}$, where $\tilde{z}_{m_ki}$ is drawn from $\mathcal{CN}(0,\tilde{\psi}_{m_ki})$ and $\tilde{\psi}_{m_k}\triangleq\sum_{i=1}^K\tilde{\psi}_{m_ki}=\mathbf{v}^{\dagger}_{m_k}\mathbf{\tilde{J}}_{m_k}\mathbf{v}_{m_k}$. As in \eqref{eq:augsoftestervar}, error matrix $\mathbf{\tilde{J}}_{m_k}$ is given as follows:
\begin{equation}\label{eq:augsoftestervar2}
\begin{split}
  \mathbf{\tilde{J}}_{m_k}=\sum_{i=1}^K\mathbf{\tilde{R}}_{mi}\oplus\tilde{\psi}_{c_k}
    \end{split}
\end{equation}
Calculation of the variance $\sigma^2_{e^{s},m_k}$ is derived in the Appendix.

\section {The CnF Strategy}
As discussed in the previous Section, when APU $m$ receives $\left\{\bar{s}_{c_k},\delta_{c_k}:k\in[1,K]\right\}$ from APU $c_k$, it creates local augmented soft estimates $\hat{s}_{m_k}$, $\forall k\in[1,K]$. Then, it needs to compare the corresponding $\text{SINR}_{m_k}$ and $\text{SINR}^{'}_{c_k}$ of $\hat{s}_{m_k}$ and $\bar{s}_{c_k}$ respectively\footnote{Note that the accent on the SINR indicates that the whole soft estimate this ratio expresses has been subjected to quantization, meaning that it carries at least one compression error, hence it is represented with a bar.}, in order to decide which estimates to transmit. 
However, APU $m$, a-priori knows that if it chooses to serve user $k$ by transmitting $\left\{\hat{s}_{m_k},\delta_{m_k}\right\}$, its contribution to them will degrade due to the digitalization process. Hence, to be more fair, it pre-distorts $\hat{s}_{m_k}$ with a random error $e_{\hat{s}_{m_k}}$\footnote{As compression errors are i.i.d. RVs, they are not known to APU $m$. However, APU $m$ does not need to calculate them. It only needs to estimate their impact on SINR, which only makes use of their variance.} and then compares it to $\bar{s}_{c_k}$. Therefore, it also takes into account the impact that this error will provoke on $\hat{s}_{m_k}$ upon its transmission to APU $m+1$, making a more realistic evaluation over its actual contribution to message $q_k$ estimation. Thus, the just comparison that needs to take place is between the SINRs of $\bar{s}_{c_k}$ and $\hat{s}_{m_k}+e_{\hat{s}_{m_k}}$, which, by applying \cite[Theorem 4.1]{emil} to \eqref{eq:ckaugsoftest}, are given as: 
\begin{equation}\label{eq:pesinrc}
  \text{SINR}^{'}_{c_k}=\frac{\rho\vert\hat{z}_{c_kk}\vert^2}{\rho\sum_{i\neq k}^K\vert\hat{z}_{c_ki}\vert^2+\rho\tilde{\psi}_{c_k}+\sigma^2_{c_k}}
 \end{equation}

\begin{equation}\label{eq:pesinr}
  \text{SINR}^{'}_{m_k}=\frac{\rho\vert\hat{z}_{m_kk}\vert^2}{\rho\sum_{i\neq k}^K\vert\hat{z}_{m_ki}\vert^2+\rho\tilde{\psi}_{m_k}+\sigma^2_{m_k}}
\end{equation}
where $\sigma^2_{c_k}=\sigma^2_n+\sigma^2_{e,\bar{y}}+\sigma^{2}_{e^{s},c_k}+\sigma^{2}_{e,\hat{s}_{c_k}}$ and $\sigma^2_{m_k}=\sigma^2_n+\sigma^2_{e,\bar{y}}+\sigma^{2}_{e^{s},m_k}+\sigma^{2}_{e,\hat{s}_{m_k}}$. Afterwards, $\forall k\in[1,K]$, APU $m$ compares \eqref{eq:pesinrc} to \eqref{eq:pesinr}, choosing the highest value, which is then marked as $\text{SINR}^{max}_{m_k}$. If $\text{SINR}^{'}_{m_k}\equiv\text{SINR}^{max}_{m_k}$, APU $m$ transmits to APU $m+1$ $\{\hat{s}_{m_k},\delta_{m_k}\}$. Otherwise, it forwards $\{\bar{s}_{c_k},\delta_{c_k}\}$.

In the final step, since APU $M$ is the CPU, no additional quantization process takes place and thus $\hat{s}_{M_k}$ is always chosen for $q_k$ detection. Hence, $\text{SINR}_{M_k}$ is calculated as in \eqref{eq:pesinrc} and \eqref{eq:pesinr}, using, though, the total noise $\sigma^2_{M_k}$, which in this case equals $\sigma^2_{M_k}=\sigma^2_n+\sigma^2_{e,\bar{y}}+\sigma^2_{e^{s},M_k}$. Using similar arguments as above, it can be shown that there is no point to compare $\text{SINR}^{'}_{u_k}$ with $\text{SINR}_{M_k}$, where $u_k$ is the APU that has lastly modified user's $k$ data before they are received from APU $M$, since $\text{SINR}^{'}_{u_k}\leq\text{SINR}_{M_k}$, $\forall k\in[1,K]$. 
Finally, the achievable uplink SE of UE $k$ can be defined as \cite[Theorem 4.1]{emil}:
\begin{equation}\label{eq:se}
\text{SE}_k=\left(\frac{\tau_c-\tau_p}{\tau_c}\right)\mathbb{E}\left\{\log_2\left(1+\text{SINR}_{M_k}\right)\right\}
\end{equation}
where the expectation is with respect to the effective channel estimates.
\section{Combining Vectors}
Following \cite[Corollary 4.3]{emil}, the optimal combining vector, for each APU $m$, that not only minimizes MSE $\{|\sqrt\rho q_k-\hat{s}_{m_k}|^2:\forall k\in[1,K]\}$, but also does not amplify unwanted error and noise terms\cite{stripes} in each step, is defined as:
\begin{equation}\label{eq:augcombvec}
\mathbf{v}_{m_k}=\frac{\mathbf{A}^{-1}_{m_k}\mathbf{\hat{h}}_{m_ki}}{||\mathbf{A}^{-1}_{m_k}\mathbf{\hat{h}}_{m_ki}||}
\end{equation}
where in case of $m=1$, $\mathbf{\hat{h}}_{m_ki}\equiv\mathbf{\hat{g}}_{1k}$, $\mathbf{v}_{m_k}\in\mathbb{C}^{L}$ and $\mathbf{A}_{1_k}\in\mathbb{C}^{L\times L}$ is given as follows:
\begin{equation}\label{eq:A}
\mathbf{A}_{1_k}=\mathbb{E}\{\mathbf{\bar{y}}_1\mathbf{\bar{y}}^{\dagger}_1\}=\sum_{i=1}^{K}\rho(\mathbf{\hat{g}}_{1i}\mathbf{\hat{g}}^{\dagger}_{1i}+\mathbf{\tilde{R}}_{1i})+\mathbf{I}_L(\sigma^2_n+\sigma^2_{e,\bar{y}})
\end{equation}
In the general case where $m>1$, $\mathbf{\hat{h}}_{m_ki}\equiv\begin{bmatrix}\mathbf{\hat{g}}_{mk} \\ \hat{z}_{c_ki} \end{bmatrix}$, $\mathbf{v}_{m_k}\in\mathbb{C}^{L+1}$ and $\mathbf{A}_{m_k}\in\mathbb{C}^{(L+1)\times(L+1)}$ can be expressed as\cite{stripes}:
\begin{align}\label{eq:augA}
&\mathbf{A}_{m_k}=\mathbb{E}\left\{\begin{bmatrix}\mathbf{\bar{y}}_m \\ \bar{s}_{c_k} \end{bmatrix}\begin{bmatrix}\mathbf{\bar{y}}_m \\ \bar{s}_{c_k} \end{bmatrix}^{\dagger}\right\}=\rho\sum_{i=1}^K\mathbf{\hat{h}}_{m_ki}\mathbf{\hat{h}}^{\dagger}_{m_ki}+\rho\mathbf{\tilde{J}}_{m_k}
\nonumber \\
+&\mathbf{I}_{L+1}(\sigma^2_n+\sigma^2_{e,\bar{y}})+\left[\mathbf{O}_L\oplus(\sigma^2_{e^{s},c_k}+\sigma^2_{e,\hat{s}_{c_k}})\right]
\end{align}
The proof for the form of the last term can be found in the Appendix.

\section{Quantization Theory Analysis}
As mentioned in previous Sections, compression of a signal $f$ produces a quantization error $e_f$, which is an i.i.d. RV assumed to be drawn from a Gaussian distribution, with zero mean and variance $\sigma^2_{e_f}$. That variance is defined \cite[para 2.5.3]{sklar}, \cite{limitedback} as:
\begin{equation}\label{eq:fvar1}
\sigma^2_{e,f}=\frac{w^2\sigma^2_f}{3\times 2^{2\alpha}}
\end{equation}
where $\sigma^2_f$ is the variance of the quantized signal, $\alpha$ is the number of bits that each symbol is converted to and $w$ is assumed to be unitary. Then, for $f=\mathbf{Y}_{p,m}$, average $\sigma^2_f$ over every AP $l$ is obtained as:
\begin{equation}\label{eq:pilotvar}
\sigma^2_{Y_{p,m}}=\mathbb{E}\left\{\left|\mathbf{Y}_{p,m}\right|^2\right\}\overset{(a)}{=}L\left(\frac{\rho\tau_p}{L}\sum_{i=1}^K tr(\mathbf{R}_{mi})+\sigma^2_n\right)
\end{equation}
where (a) follows from the independence between user channels. In case $f=\mathbf{y}_m$, average $\sigma^2_f$ over AP $l$ and APU $m$ is calculated as:
\begin{equation}\label{eq:datavar}
\sigma^2_{\bar{y}}=\mathbb{E}\left\{\left|\mathbf{y}_{m}\right|^2\right\}\overset{(a)}{=}L\left(\frac{\rho}{LM}\sum_{m=1}^M\sum_{i=1}^Ktr(\mathbf{R}_{mi})+\sigma^2_n\right)
\end{equation}
where the expectations in \eqref{eq:pilotvar} and \eqref{eq:datavar} are over channel $\mathbf{g}_{mi}$. Also, using random vector $\mathbf{\hat{h}}_{m_ki}$ as defined in previous Section, $\sigma^2_{\hat{s}_{m_k}}$ can be calculated as:
\begin{equation}\label{eq:svar}
\begin{split}
\sigma^2_{\hat{s}_{m_k}}&=\mathbb{E}\left\{|\hat{s}_{m_k}|^2\vert\{\mathbf{\hat{h}}_{m_ki}\}\right\}=\mathbf{v}^{\dagger}_{m_k}\mathbf{A}_{m_k}\mathbf{v}_{m_k}
\end{split}
\end{equation}

\begin{table}[h]
\centering
\begin{tabular}{|c||c|c|}
\hline
\textbf{Component} & \textbf{Signal transmitted} & \textbf{Total scalars}\\
\hline \hline
\multirow{3}{*}{APU to APU} & $\hat{s}_{m_k}\in\mathbb{C}^{1\times 1},\forall k\in[1,K]$ & $K\tau_d$ \\
\cline{2-3}
&$\hat{z}_{m_ki}\in\mathbb{C}^{1\times 1},\forall k,i\in[1,K]$ & $K^2$\\
\cline{2-3}
&$\tilde{\psi}_{m_k}\in\mathbb{C}^{1\times 1},\forall i\in[1,K]$ & $K$\\
\hline \hline
\multirow{2}{*}{APs to APU} & $y_{p,ml}\in\mathbb{C}^{1\times\tau_p},\forall l\in[1,L]$ & $(L-1)\tau_p$ \\
\cline{2-3}
& $y_{ml}\in\mathbb{C}^{1\times 1},\forall l\in[1,L]$ & $(L-1)\tau_d$ \\
\hline
\end{tabular}
\caption{Traffic data between successive APUs in each coherence interval $\tau_c=\tau_d+\tau_p$.}
\label{table:1}
\end{table}

Then, in order to calculate the minimum bit rate needed in order to have an error-free connection, one should focus on the amount of complex scalars every component of the network needs to transmit in each $\tau_c$. These are summarized in Table \ref{table:1}. According to these data and for a fixed radio stripe capacity $C$, in order to have a lossless connection, maximum resolutions $\alpha_{AP}$ and $\alpha_{APU}$, appearing in \eqref{eq:fvar1}, that the quantizers of APs and APUs respectively possess, should satisfy the following:
\begin{equation}\label{eq:alpha}
\begin{split}
& \alpha^{cf}_{AP}=\frac{r CT_c}{2(L-1)\tau_c} \left(\frac{bits}{symbol}\right)\\
& \alpha^{cf}_{APU}= \frac{(1-r)CT_c}{2K(\tau_c+1)} \left(\frac{bits}{symbol}\right)
\end{split}
\end{equation}
where $T_c (sec)$ is the coherence time. Note that $r\in[0,1]$ indicates the capacity percentage allocated for the transmission of pilot and data signals, while $1-r$ is the capacity percentage allocated for the transmission of soft estimates and side information. Clearly, for the case of APU 1, $r=1$, since only APs need to use the FH link to transmit information \footnote{According to eq. \eqref{eq:fvar1}, in order to average $\sigma^2_{e,\bar{y}}$ over all APs of the system, we also considered average $a_{AP}$ resolution. That is only applied in $\sigma^2_{e,\bar{y}}$ calculation.}. 

\section{Numerical Results}
We consider the stripe of Fig. \ref{2b} that has its APs and APUs uniformly distributed along its length. The film is placed 4.5 m above ground and the users are equally spread lengthwise of it at a fixed distance of 5 m from it. For comparison, we also include the scheme of \cite{stripes}, which differs from ours because:
\begin{itemize}
\item APs are co-located on every APU and thus APUs occupy all the FH capacity ($r=0)$.
\item Each UE is served from all $M$ APUs.
\item APU $M$ differs from CPU, hence one additional quantization takes place before $\hat{s}_{M_k}$ reaches its final destination.
\item Inter-APU scalars are $2K(\tau_c+K)$, since each APU transmits $\tilde{\psi}_{m_ki}$, $\forall k,i\in[1,K]$. Hence, $\alpha^{cl}_{APU}$ is modified accordingly.
\end{itemize} 
Since this design is meant for outdoor applications, it matches well with the 3GPP Urban Microcell model in \cite[Table B.1.2.1-1]{3GPP}, which defines path loss as follows:
\begin{equation}\label{eq:pl}
\text{PL (dB)}=-36.7\text{log}_{10}\left(d_{mlk}\right)-26\text{log}_{10}\left(f_c\right)-22.7
\end{equation}
where $d_{mlk}$ is the direct distance between UE $k$ (assumed 1.5 m above ground) and AP $l$ of APU $m$, taking also into account the height of the stripe and its horizontal distance from the users. Carrier frequency $f_c$ is set to 2 GHz and noise power $\sigma^2_n$ to -92 dBm.
\begin{figure}[ht]
    \centering
    \includegraphics[trim={0 0 0 0},clip,width=.90\linewidth]{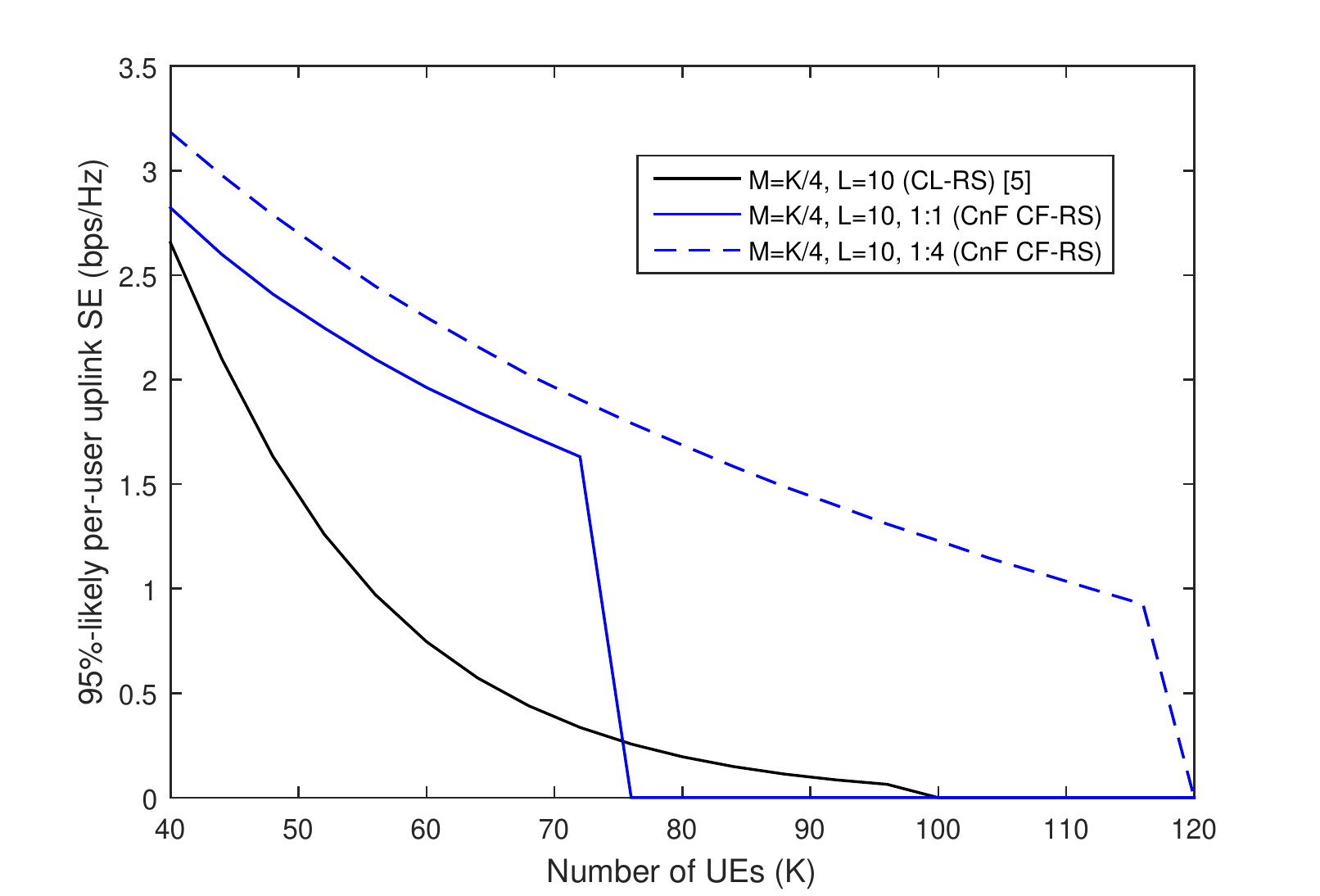}
    \caption{95\%-likely per-user uplink SE as a function of number of UEs served.}
    \label{plot1}
\end{figure}
\begin{figure}[ht]
    \centering
    \includegraphics[trim={0 0 0 0},clip,width=.90\linewidth]{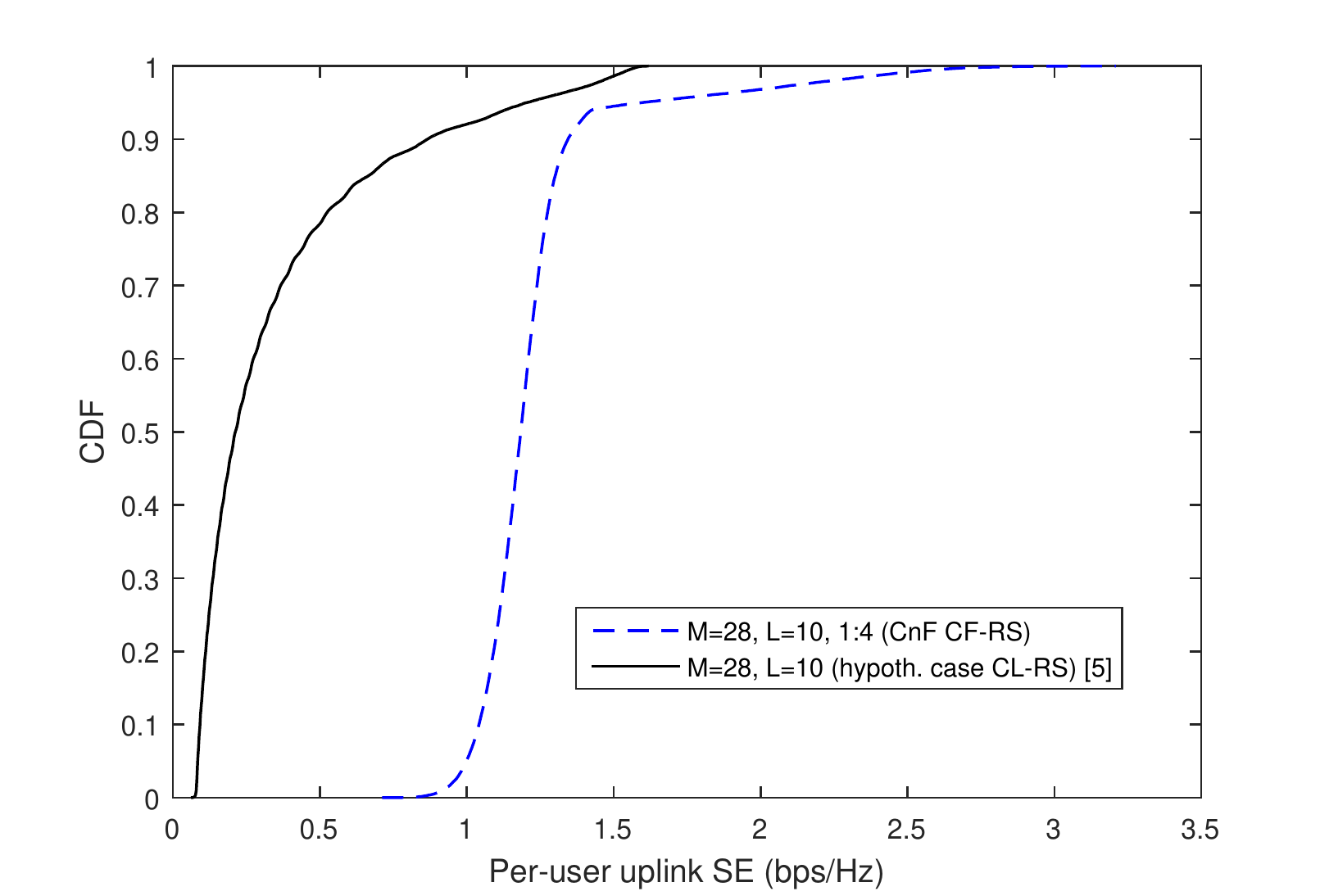}
    \caption{CDF as a function of the per-user uplink SE. Here, $\alpha^{cl}_{APU}\equiv\alpha^{cf}_{APU}$ (hypoth. case). }
    \label{plot2}
\end{figure}

In addition, we envision a medium that has coherence bandwidth $B_c=100$ kHz and coherence interval $\tau_c=180$ samples. The arrangement uses 20 MHz spectral bandwidth and the radio stripe capacity is limited to 6 Gbps (e.g. Ethernet Cat 6a), meaning that it disposes 30 Mbps for each $\tau_c$. Also, $\tau_p$ is set equal to the number of UEs $K$, transmit power $\rho$ of UEs equal to 100 mW, inter-user distance (DBU) equal to 3 m (each terminal addition extends the system by 3 m), while spacing between APUs and APs (only in our case) is properly chosen in order for the stripes to cover exactly the same extent terminal devices occupy. In Fig. \ref{plot1} we compare our arrangement, using two different capacity allocation ratios $r:(1-r)$ and the CnF strategy, to the one in \cite{stripes} in terms of 95\%-likely per-user uplink SE as number of UEs and APUs scale. The CL radio stripe appears to drop its performance at a higher pace compared to all the other schemes, which is an outcome of two main reasons. The first is that as we increase number of users $K$, more scalars need to be transferred and hence we can dedicate only a few bits for the representation of each symbol, while the second one, is caused by the increase of APUs, which impose more mandatory quantizations to take place. On the contrary, our proposition seems to handle the expansion of the system better in any case. This happens because each user $k$ is only served by APUs ($\text{DCC}_k$) that truly enhance their $\text{SINR}_k$, hence the number of quantizations does not scale as APUs do. However, performance drop as $K$ increases is inevitable. Thus, in the case where we split the capacity into two equal parts (1:1 CnF CF-RS case), link termination (less than 1 bit per symbol) occurs faster than it occurs in the case of the CL radio stripe where all the capacity is used for inter-APU data transferring. Nevertheless, since $\alpha^{cf}_{AP}$ stays constant as system scales, reallocating more rate (1:4 CnF CF-RS case) for inter-APU data transferring not only permits for higher performance, but also for more users to be served.

Fig. \ref{plot2} shows the CDF of the per-user uplink SE, using the same parameters as in the setup of Fig. \ref{plot1}, adjusted, though, to the instance of 112 UEs. However, as seen in the previous figure, the CL radio stripe cannot serve 112 UEs. Nevertheless, to eliminate the possibility that the higher performance of our proposed scheme is due to smaller quantization error variances rather than the novel user-centric DDC scheduling, we hypothetically set $\alpha^{cf}_{APU}=\alpha^{cl}_{APU}$. Even then, the results showcase that, the standard CL-RS scheme, even under theoretical error reduction, offers less throughput to 100\% of the users and 14 times less 95\%-likely per-user throughput when compared to our proposition. That is a consequence of the enhanced macro-diversity appearing in our arrangement, but most importantly due to the balance accomplishment between the contribution that each $\hat{s}_{m_k}$ offers to $q_k$ estimation and the deterioration that quantization errors provoke to these soft estimates upon their transmission into the finite-capacity radio stripe.

\section{Conclusion}
Radio stripe topologies promise to address Cell-Free Massive MIMO issues, such as high implementation costs, adaptation rigidness and computation accretion. However, limited capacity of such schemes makes their implementation almost impossible. In this paper we analysed the impact that a finite-capacity radio stripe has on the per-user uplink SE and we proposed a heuristic strategy to turn such serial topologies into user-centric networks, making them function more efficiently. By using our CnF strategy, we were able to run DDC scheduling, an action that radically helps in redundant signal compression avoidance throughout the sequential proceeding. That way we managed not only to enhance per-user uplink SE, but also to set the basis for our future work, which includes more complex, tree-like radio stripe networks. Additionally, we suggested a new radio stripe architecture, which better exploits the benefits of macro-diversity. Ultimately, as numerical results showcase, our proposition offers much better results, especially in scaled scenarios, than the existing sequential scheme of \cite{stripes}, rendering it more appealing for real-world implementations.

\section*{Appendix}
In this Appendix, we will evaluate the variance of $e^{s}_{m_k}$ defined in \eqref{eq:randrecaug}. Without loss of generality, assume that UE $k$ is served by a subset $\text{DCC}_k\subseteq\{1,2,\dots,M\}$ of $D$ APs, where $D\leq M$. That subset (cluster) is given as $\text{DCC}_k=\{d_1,d_2,\dots,d_D\}$, where $d_1=1$ and $d_D=M$ since APU 1 and M always serve UE $k$. Then, $e^{s}_{d_{2k}}$ in $\hat{s}_{d_{2k}}$ equals to:

\begin{equation}\label{eq:esd2k}
e^{s}_{d_{2k}}=\mathbf{v}^{\dagger}_{d_{2k}}\left[\begin{array}{cc}\mathbf{0}^T_L & e_{\hat{s}_{2_k}}\end{array}\right]^T 
\end{equation}
where $e^{s}_{d_{2k}}\sim\mathcal{CN}(0,\sigma^2_{e^{s},d_{2k}})$ and $\sigma^2_{e^{s},d_{2k}}$ to be given as:

\begin{equation}\label{eq:esd2kvar}
\begin{split}
\sigma^2_{e^{s},d_{2k}}=\mathbb{E}\left\{\left|e^{s}_{d_{2k}}\right|^2\right\}=\mathbf{v}^{\dagger}_{d_{2k}}\left(\mathbf{O}_{L}\oplus\sigma^2_{\hat{s}_{2_k}}\right)\mathbf{v}_{d_{2k}}
\end{split}
\end{equation}
In APU $d_3$, received $\bar{s}_{d_{2k}}$ not only burdens $\hat{s}_{d_{3k}}$ with $e^{s}_{d_{2k}}$, but also with $e_{\hat{s}_{d_{2k}}}$ due to its quantization. New $e^{s}_{d_{3k}}$ is given as:
\begin{equation}\label{eq:esd3k}
\begin{split}
e^{s}_{d_{3k}}=\mathbf{v}^{\dagger}_{d_{3k}}\left[\begin{array}{cc}\mathbf{0}^T_L & e^{s}_{d_{2k}}+e_{\hat{s}_{d_{2k}}}\end{array}\right]^T
\end{split}
\end{equation}
where $e^{s}_{d_{3k}}\sim\mathcal{CN}(0,\sigma^2_{e^{s},d_{3k}})$. Exploiting that $e^{s}_{d_{2k}}$ and $e_{\hat{s}_{d_{2k}}}$ are independent RVs, $\sigma^2_{e^{s},d_{3k}}$ is calculated as:
\begin{equation}\label{eq:esd3kvar}
\sigma^2_{e^{s},d_{3k}}=\mathbb{E}\left\{\left|e^{s}_{d_{3k}}\right|^2\right\}
=\mathbf{v}^{\dagger}_{d_{3k}}\left[\mathbf{O}_{L}\oplus(\sigma^2_{e^{s},d_{2k}}+\sigma^2_{e,\hat{s}_{d_{2k}}})\right]\mathbf{v}_{d_{3k}}
\end{equation}
where \eqref{eq:esd3kvar} can be calculated using \eqref{eq:esd2kvar} and \eqref{eq:svar}.

In a similar way, general cumulative quantization error variance $\sigma^2_{e^{s},m_k}$ can be expressed as:
\begin{equation}\label{eq:esmkvar}
\begin{split}
\sigma^2_{e^{s},m_k}&=\mathbf{v}^{\dagger}_{m_k}\left[\mathbf{O}_{L}\oplus(\sigma^2_{e^{s},c_k}+\sigma^2_{e,\hat{s}_{c_k}})\right]\mathbf{v}_{m_k}
\end{split}
\end{equation}

\bibliographystyle{IEEEtran}
\bibliography{IEEEfull,references}

\end{document}